\documentstyle[gsirep,epsfig]{article}
\title{Excitation function of entropy and pion production from
AGS to SPS energies}
\author{M. Reiter, A. Dumitru, J. Brachmann, J.A. Maruhn, H. St\"ocker,
W. Greiner\\
Institut f\"ur Theoretische Physik, Universit\"at Frankfurt a.M., Germany}
\date{}
\begin{document}
\maketitle

\noindent 
In the initial off-equilibrium stage of energetic nuclear collisions
a large amount of entropy can be produced by nuclear shockwaves \cite{entropy},
while the subsequent expansion is often assumed to be nearly isentropic.
In such a scenario the entropy produced during compression is closely
linked to the finally observed particle yields.
In particular, it has recently been
noted \cite{Pions} that in heavy-ion collisions at AGS energy less pions
per participating baryon are produced than in proton-proton reactions
at the same energy (per nucleon). At the higher SPS energy, however, this
difference was found to be positive: the number of pions per baryon in 
$S+S$ is larger than in $p+p$ at $\sqrt{s}\approx20~GeV$. 

We have calculated entropy production in the compression stage of
heavy-ion collisions within the 3-fluid dynamical model \cite{3f}.
A possible increase of specific entropy during expansion due to finite
viscosity has been neglected.
Within this model, the particles involved in a reaction are divided
into three separate fluids: the first two fluids correspond to
the projectile and target nucleons, respectively,
and the particles produced during the reaction are collected in the third
fluid. Local thermodynamic equilibrium is maintained only in each fluid
separately but not between the fluids. The fluids are able to penetrate
and decelerate each other during the collision. 
Interactions between projectile- and target-fluid are due to
binary collisions of the nucleons. This allows for a treatment of
non-equilibrium effects in the initial stage of the collision.
In particular, due to the finite mean free path, the entropy-generating shock
fronts are smeared out considerably \cite{3f}, in contrast to (ideal) one-fluid
hydrodynamics, where they are sharp discontinuities.
In the present calculation, we have employed an EoS (for all three fluids)
with first order phase transition to QGP. Further details of the calculation
will be presented elsewhere.

\begin{figure}
\vspace*{-.5cm}
\centerline{\hbox{\psfig{figure=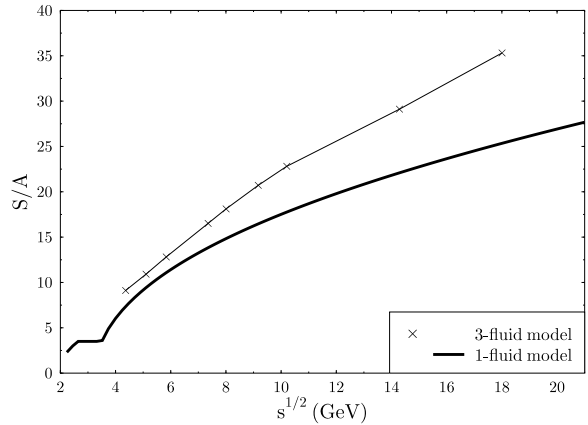,width=10cm}}}
\vspace*{-.5cm}
\caption{Entropy produced in the initial compression stage per
net baryon as a function of beam energy.}
\label{fig1}
\end{figure}

The excitation function of the entropy per participating net-baryon as calculated
in both the 3-fluid model and the 1-dimensional 1-fluid shock model is depicted
in Fig.~\ref{fig1}. One observes that both $S/A$-excitation functions are 
continuous functions 
of $\sqrt{s}$ and do not exhibit a jump at some specific energy. 
In the 1-fluid shock model a plateau is observed
\cite{rigo}, which is due to the disappearance of the sharp single compression
shock wave \cite{hof}. Due to the smaller inelasticity in the 3-fluid model, 
this plateau is expected to shift to higher energies, $\sqrt{s}\approx 
\mbox{5-10}$~AGeV.
However, the broadened shock fronts lead to a smooth increase of entropy 
production instead of a sharp threshold behaviour. Also, the ratio of thermal
 to compression
energy is higher within the 3-fluid model, leading to increased entropy per 
baryon at all energies (e.g.\ $S/A=35$ to 25 at SPS).

\begin{figure}
\vspace*{-.5cm}
\centerline{\hbox{\psfig{figure=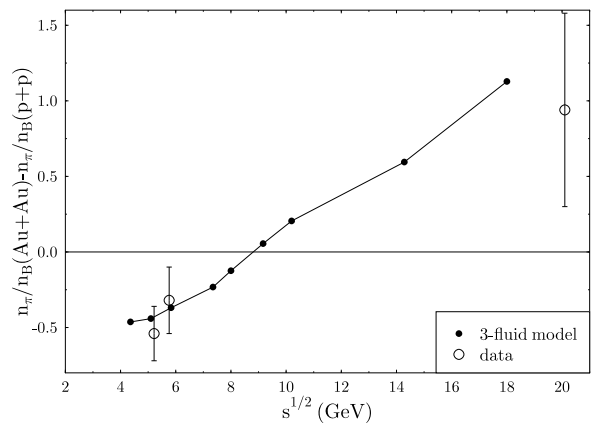,width=10cm}}}
\vspace*{-.5cm}
\caption{
Number of pions per participating baryon in $Au+Au$ minus that in $p+p$
(feeding from resonance decay is taken into account).}
\label{fig2}
\end{figure}

Assuming entropy conservation during expansion and a freeze-out density of 
$\rho_{\mbox{fo}}= 0.5 \rho_0$ (for all energies), pion to baryon ratios as shown in 
Fig.~\ref{fig2} are 
obtained. These are in good agreement with experimental data \cite{Pions}.
The sign reversal of $\Delta(n_\pi/n_B)$ between AGS and SPS can
be understood as being due to excess ``non-baryonic'' entropy
produced at SPS.
Thus, the smooth increase of $S/A$ reflects in a continuous increase of 
$\Delta(n_\pi/n_B)$ with $\sqrt{s}$.

\end{document}